\documentclass{article}
\usepackage{graphicx}
\usepackage[utf8]{inputenc}
\usepackage{amsmath}
\usepackage{comment}
\usepackage[letter paper, total={6.5in, 9in}]{geometry}

\title{Kinetics of Bose-Einstein condensation of magnons\\ in Yttrium Iron Garnet films}
\author{$^{1^*}$Hulin Yang, $^{1}$Gang Li, $^{2}$Haichen Jia, $^{1}$Artem Abanov, $^{1,3}$Valery Pokrovsky \\[0.5em]
$^{1}$Physics Department, Texas A\&M University, College Station, Texas, USA \\
$^{2}$Physics Department, The University of Hong Kong, Hong Kong \\
$^{3}$Landau Institute for Theoretical Physics, Chernogolovka, Moscow District, Russia}
\date{}
\begin{document}
\maketitle

\section*{Abstract}
In this article, we explain the reason of the apparent contradiction between recent experiments \cite{Borisenko1} and \cite{Borisenko2} and earlier theoretical predictions \cite{Fuxiang-Li} of strongly asymmetric condensate resulting in attractive interaction between the condensate magnons.
We show that the relaxation time for equilibrium between two condensates at two minima of energy exceeds the time of experiment. Therefore, it should be described by Boltzmann kinetic equation. We develop the proper kinetic theory and find the relation between the critical pumping power and the effective temperature of over-condensate magnons.

\section{Introduction}

Let $J$ be the exchange interaction energy between nearest-neighbor spins, $a$ the lattice constant, and $z$ the coordination number. The characteristic dipolar interaction energy is 
$E_{\text{dip}} = \frac{2\mu_B^2}{a^3}$, where $\mu_B$ is the Bohr magneton.
The so-called dipolar length is defined as $\ell_{\text{dip}} = \frac{Jza}{E_{\text{dip}}}$, it serves as a length scale at which the dominance of interactions changes: at distances smaller than $ \ell_{\text{dip}} $, the dipolar interaction dominates, while at larger distances, the exchange interaction prevails.
In YIG at room temperature, the dipolar length is approximately 40 nm. It has been reliably established \cite{Sonin, Gang-Li} that the spectrum of magnons in a thick ferromagnetic films with thickness $ d \gg \ell_{\text{dip}} $ has two minima at wave vectors $\mathbf{k}= \pm \mathbf{Q}$ collinear with spontaneous magnetization. Let us choose  $z$ -axis parallel to the spontaneous magnetization $\mathbf{M}$, the coordinate $y$ be parallel to the surface and perpendicular to $z$ and the direction perpendicular to the surfaces of the film be $x$-axis. The spectrum of magnons, i.e., their frequencies $\omega_n\left( k_y,k_z\right)$ are characterized by the number of the branch (standing wave in $x$-direction) $n$ and wave-vector $\mathbf{k}=\left(k_y,k_z\right)$ in the surface plane. Magnetic field $\mathbf{H}$ fixes the direction of spontaneous magnetization and creates the Zeemann gap in the spectrum, so that the frequency at the two minima is $\omega_0(\pm \mathbf{Q}) = \gamma H$, where $\gamma = \frac{e}{mc} $ is doubled gyromagnetic ratio.
The maximum frequency at $\mathbf{k} = 0$ corresponds to the ferromagnetic resonance: $\omega(0) = \gamma \sqrt{H(H + 4\pi M)}$, where  $M$  is the magnetization
.Below we show the exact expression for the spectrum of magnons in the plane $k_y=0$.
\begin{equation}\label{eq:spectrum z-x}
\omega_n(k_z) = \gamma\sqrt{H^2+2HM\ell^2\left( k_{xn}^2+k_z^2\right)+\frac{4\pi HM k_{xn}^2}{\left( k_{xn}^2+k_z^2\right)}}
\end{equation}
The values $k_{xn}$ satisfies equation $tan\frac{k_{xn}d}{2}=k/k_x$ or  $tan\frac{k_{xn}d}{2}=-k_x/k$, where $k=\sqrt{k_z^2+k_{xn}^2}$. Thus there are two series of $k_{xn}$ that will be denoted by an upper index $\pm$. Asymptotically at large $n$, $k_{xn}^{\pm}=(2\pi n - \pm \pi/2)/d$.

In the year 2006 S. O. Demokritov et al. \cite{Demokritov} discovered that in thick film of YIG, there appears the Bose-Einstein condensate of magnons (BECM) at room temperature when the pumping power $P$ exceeds some critical value.Recent studies demonstrated that the Dzyaloshinskii-Moriya interaction (DMI) renormalizes the magnon spectrum and enhances the separation of the two energy minima, thereby favoring the coexistence of distinct condensates \cite{Kemayou}. Nonequilibrium condensation phenomena are not unique to magnons: similar driven-dissipative Bose-Einstein condensation has been widely studied in photonic systems, revealing universal features of coherence generation and nonequilibrium critical behavior \cite{Carusotto}. This broader context highlights the importance of considering nonequilibrium dynamics in magnon condensates.

Magnon Bose-Einstein condensation and spin superfluidity have been demonstrated experimentally and theoretically in diverse systems, from superfluid He 3-B to magnetic
materials such as YIG \cite{Borovik-Romanov, Bunkov, Novik-Boltyk, ChenSun, Serga, Bozhko}. In recent years, numerous studies have explored related nonequilibrium condensation phenomena, nonlinear magnon interactions, and hybrid excitations across a broad range of physical platforms \cite{Koster, ZhangZ, Bloch, Hurst, ZhangJ, Xu, Mohseni, Shan}. These works demonstrate the growing interest and rapid progress in the field. However, a comprehensive statistical description of the BECM dynamics based on first-principles calculations is still lacking.

\section{Quasi-equilibrium state}
Without external pumping, the occupation numbers of magnons obey the Bose–Einstein statistics,
\begin{equation}\label{eq:occupation-numbers}
n(m,\mathbf{k})=\frac{1}{e^{\frac{\epsilon_m(\mathbf{k})}{k_B T}}-1}
\end{equation}
The magnon relaxation time depends on energy. For thermal magnons with energy $\sim k_B T$, the relaxation time is about 10 ps. However, at low energy the relaxation time is much longer. 

In the process of external pumping, a photon with frequency $\omega_0$ generates two magnons with almost opposite wave vectors $\pm \mathbf{k}$ and frequency $\omega_m(\mathbf{k})=\omega_0/2$. If $\omega_0$ is chosen so that $\omega_0/2<2\gamma H$, the decay process is forbidden. Still the processes of merging of a low energy magnon with a thermal magnon or inverse process of Cherenkov radiation of low energy magnon by a thermal magnon are possible reducing the lifetime of low-energy magnons. But the matrix elements of these processes are strongly suppressed by very small Mach angles. Thus, one can expect that the relaxation time $\tau_r$ for low energy magnons is still much shorter than their lifetime $\tau_l$. It means that during the relaxation process, the number of magnons is conserved. Therefore they reach the equilibrium state with a finite chemical potential $\mu$. 

In a larger time scale, the newly pumped magnons relax and balance the decayed magnons. In other words, after the pumping is turned on, the system asymptotically approaches a new stationary and equilibrium state characterized by a chemical potential $\mu$. Stronger the pumping power is, the bigger is the chemical potential. In the quasi-equilibrium state, the occupation number of a state with fixed number of branch $m$ and wave vector $\mathbf{k}$ is
\begin{equation}\label{eq:BE-mu}
n(m,\mathbf{k})=\frac{1}{e^{\frac{\epsilon(m,\mathbf{k})-\mu}{k_BT}}-1}
\end{equation}
At pumping power so large that the chemical potential becomes equal to the gap in the spectrum $\Delta=\epsilon_0(\mathbf{Q})=\hbar\gamma H$, the occupation number (\ref{eq:BE-mu}) for the first branch $m=1$ becomes infinite. At larger pumping power magnons start to condensate.
Such assumption was accepted in the work \cite{Fuxiang-Li} which concluded that in the quasi-equilibrium state the magnons occupy the two minima in a very asymmetric way that leads to effective attraction between magnons. However, the recent experimental work by the S. O. Demokritov's team \textit{et al.} \cite{Borisenko1} showed that magnons repulse each other. 

\begin{figure}[!ht]
    \centering
    \includegraphics[width=0.5\textwidth]{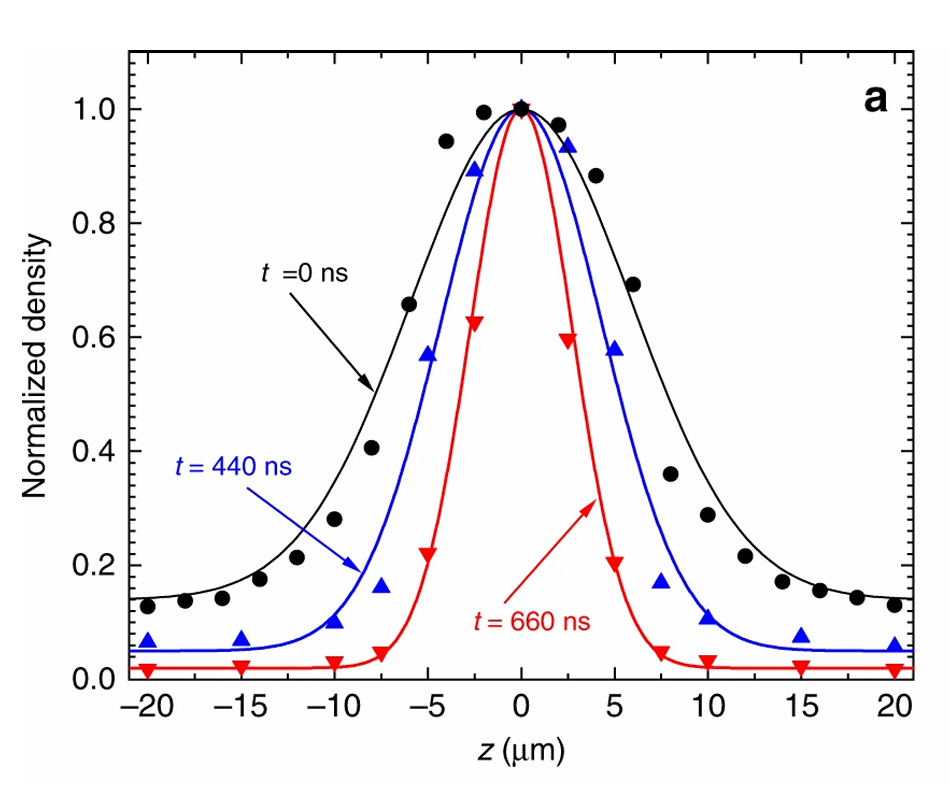}
    \caption{Time evolution of the normalized condensate density after switching off the pumping. The density in the graph is normalized with respect to the initial density of condensate magnons. Without pumping the magnon density decreases exponentially in time. Each curve show the distribution of magnon density at a fixed time. The width of curves contracts with time or equivalently with the density that proves the repulsive interaction. Reprinted from \cite{Borisenko1}. Used under https://creativecommons.org/licenses/by/4.0/.}
    \label{repulsive_norm_density}
\end{figure}

In this experiment they switched off the pumping and measured the dependence of magnetization on coordinate  at different moments of time. The corresponding curves normalized to 1 at maximum are shown in Fig.\ref{repulsive_norm_density}.  Since without pumping the magnetization decreases exponentially with time, it can serve as rescaled average density. Looking at Fig.\ref{repulsive_norm_density}, we observe that the average distance between magnons is the larger the larger is their density. It means that magnons repulse each other.

\begin{figure}[!ht]
    \centering
    \includegraphics[width=0.5\textwidth]{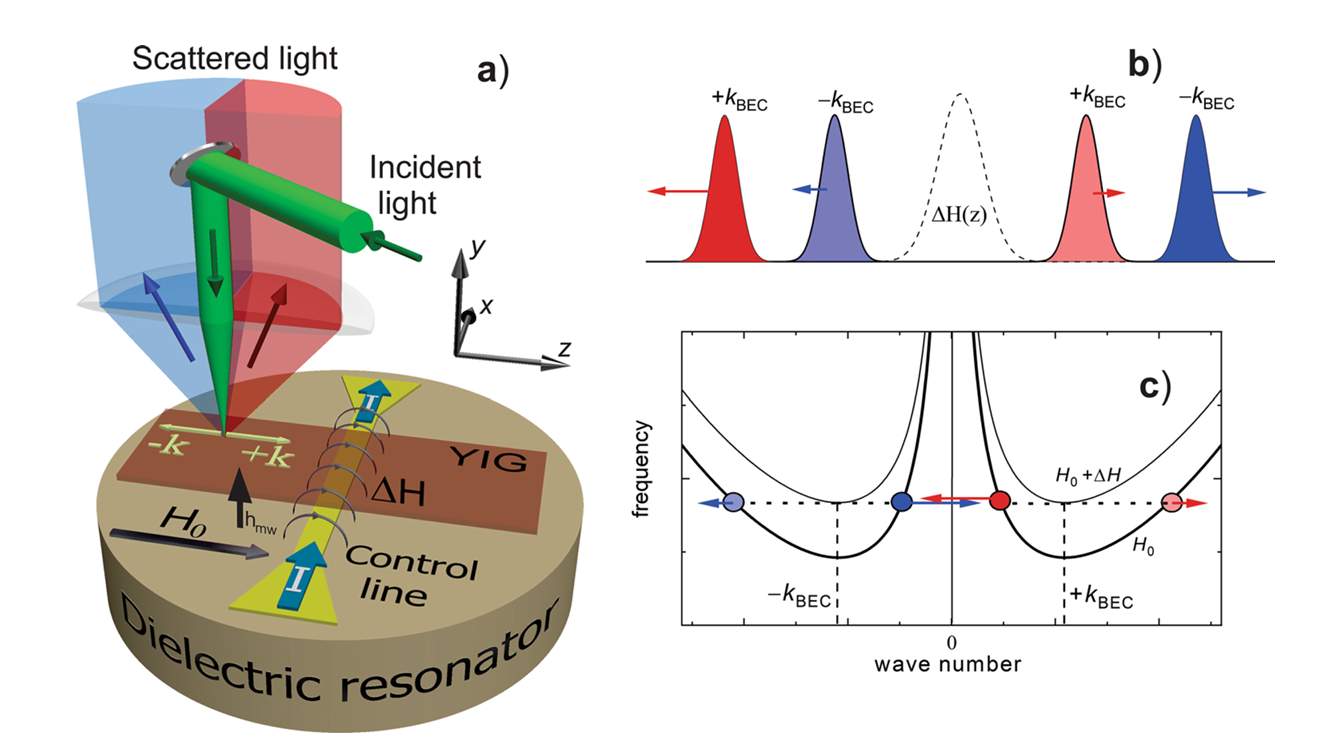}
    \caption{ (a) A schematic representation of the experimental setup used in the study. (b) Following the application of a localized magnetic field pulse, the Bose-Einstein condensate (BEC) was observed to split into four distinct sub-clouds. Each sub-cloud propagated in a specific direction, determined by the condensate's initial wave vector and the slope of the relevant magnon dispersion branch. (c) The magnon dispersion relations were examined at two spatial locations: one at the center of the metal strip, where the magnetic field strength was $H_0+\Delta H$ and another at a more distant point with field strength $H_0$. Colored markers denote the spectral positions of the magnon states corresponding to the four sub-clouds after their separation. The associated group velocities, and thus their directions of motion in real space, are indicated schematically with arrows. Reprinted from \cite{Borisenko1}. Used under
https://creativecommons.org/licenses/by/4.0/.}
    \label{separation}
\end{figure}

In the second experiment \cite{Borisenko2}, the experimenters submitted to condensate an initial pulse of energy. As a result, instead of staying at $k=\pm Q$, the two condensates separated into four, two of them moving to the right and two others to the left, as it is shown in Fig.\ref{separation}. Though they are asymmetric, this asymmetry is not so dramatically strong as predicted by theory \cite{Fuxiang-Li} and may be not intrinsic, but generated by asymmetry of the experimental device.

It shows that a  basic  assumption of the theory \cite{Fuxiang-Li} is invalid. Since the magnons Hamiltonian (the spectrum) was checked by independent experiment, the only assumption that can be invalid is the quasi-equilibrium assumption. In other words, it means that  there exists a degree of freedom whose relaxation time exceeds the time of the experiment. We already know that relaxation time inside one minimum is few tens nanoseconds. It is governed by the process of scattering of two magnons into two. But these processes do not contribute to the relaxation of inter-minima degree of freedom. In the next section we show that indeed the inter-minima relaxation time is very long.

\section{Inter-minima relaxation time}
\begin{figure}[!ht]
    \centering
    \includegraphics[width=0.5\textwidth]{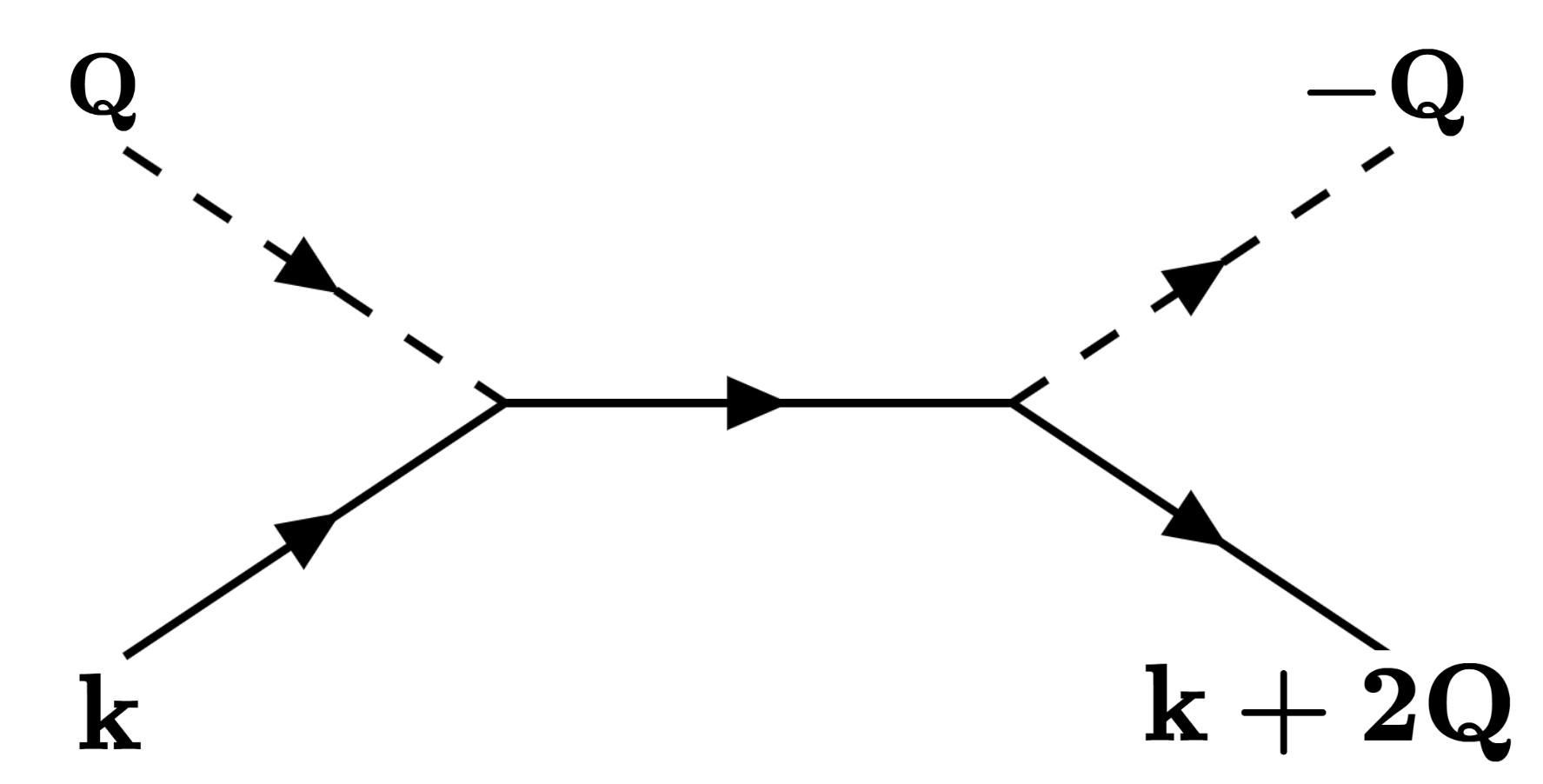}
    \caption{Compton scattering, consists of two consecutive 3-magnon processes. The scattering amplitude is proportional to the product of two 3rd order Hamiltonian verticis. Using the Fermi's golden rule for the inter-minima transition rate, we find the relaxation time $10^6\,s$, which is much longer than the lifetime of the condensate magnons, indicating that the condensate magnons are under-relaxed, not reaching their thermodynamic equilibrium.}
    \label{compton}
\end{figure}

\begin{figure}[!ht]
    \centering
    \includegraphics[width=0.5\textwidth]{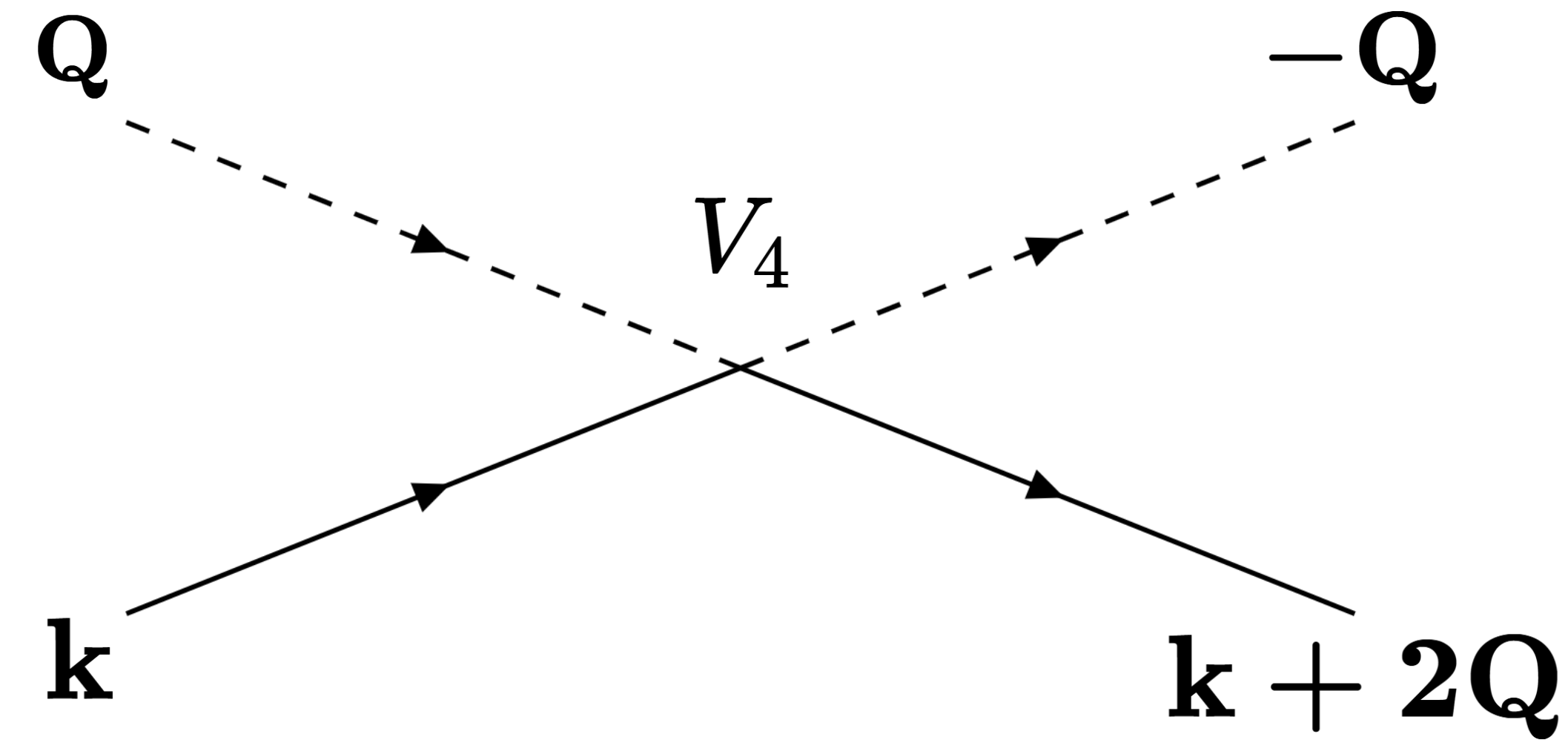}
    \caption{Direct 4-magnon processes, with amplitude proportional to the 4th order vertex. Using the Fermi's golden rule, we find the relaxation time $\sim 10^5\,\mathrm{s}$, which is also much longer than the lifetime of the condensate magnons and time of the experiment.}
    \label{V4}
\end{figure}

Two channels contribute to the inter-minima relaxation: Compton scattering shown in Fig.\ref{compton}, and 4-vertex process shown in Fig.\ref{V4}. In each of these two processes a low energy magnon with momentum $\mathbf{Q}$ is absorbed by a high-energy magnon with momentum $\mathbf{k}$ and by Cherenkov radiation is transferred to the second minimum $-\mathbf{Q}$.

The decay and merging processes in Compton scattering are described by the vertex $V\left(\mathbf{k}, k_x ; \mathbf{k}^{\prime}, k_x^{\prime}\right)$, that is completely due to the dipolar interaction. It can be represented as
\begin{equation}\label{eq:dip-vert}
   V\left(\mathbf{k}, k_x ; \mathbf{k}^{\prime}, k_x^{\prime}\right)=\frac{8 \pi^2\left(2 \mu_B\right)^{3 / 2} M^{1 / 2}}{i \sqrt{A d}} \frac{k_y k_z}{\mathbf{k}^2+k_x^2}\left[\frac{\cos \frac{\left(k_x-k_x^{\prime}\right) d}{2}}{\left(k_x-k_x^{\prime}\right)^2 d^2-\pi^2}+\frac{\cos \frac{\left(k_x+k_x^{\prime}\right) d}{2}}{\left(k_x+k_x^{\prime}\right)^2 d^2-\pi^2}\right] ,
\end{equation}
where $A$ is the area and $d$ is the thickness of the film. Employing this vertex, the amplitude of the Compton process is
\begin{equation}\label{eq:Compton}
\mathcal{A}=d \int_{-\infty}^{\infty} \frac{d k_{1 x}}{(2 \pi)} \frac{V_1 V_2^* \sqrt{N_{\mathbf{k}+\mathbf{Q}, k_{1 \mathrm{x}}}\left(N_{\mathbf{k}+\mathbf{Q}, k_{1 \mathrm{x}}}+1\right)}}{\epsilon\left(\mathbf{k}, k_x\right)+\epsilon(\mathbf{Q})-\epsilon\left(\mathbf{k}+\mathbf{Q}, k_{1 \mathrm{x}}\right)+i \delta},
\end{equation}
where $V_1=V\left(\mathbf{k}, k_x, \mathbf{k}_1, k_{1 x}\right), V_2=V\left(\mathbf{k}+2 \mathbf{Q}, k_x^{\prime}, \mathbf{k}_1, k_{1 x}\right)$ The inverse relaxation time or probability of the Compton process per unit time $1 / \tau_C$ is given by:
\begin{equation}\label{eq: tau-C}
\frac{1}{\tau_C}=\frac{2 \pi\left|N_{\mathbf{Q}}-N_{-\mathbf{Q}}\right|}{\hbar} A d^2 \int \frac{d^3 k d k_x^{\prime}}{(2 \pi)^4}|\mathcal{A}|^2 N_{\mathbf{k}, k_x}\left(N_{\mathbf{k}, k_x}+1\right) \delta\left[\epsilon\left(\mathbf{k}, k_x\right)-\epsilon\left(\mathbf{k}+2 \mathbf{Q}, k_x^{\prime}\right)\right]
\end{equation}
Since $|\mathcal{A}|^2=(\Im \mathcal{A})^2+(\Re \mathcal{A})^2$, let us first estimate $\Im \mathcal{A}$ :
\begin{equation}\label{eq: Im A}
\Im (\mathcal{A})=\frac{d}{2} \int_{-\infty}^{\infty} V_1 V_2^* \sqrt{N_{\mathbf{k}+\mathbf{Q}, k_{1 x}}\left(N_{\mathbf{k}+\mathbf{Q}, k_{1 x}}+1\right)} \delta\left(\epsilon_{\mathbf{k}+\mathbf{Q}, k_{1 x}}-\epsilon_{\mathbf{k}, k_x}\right) d k_{1 x} .
\end{equation}
The energy of a thermal magnon is almost completely determined by exchange interaction. Therefore, it is proportional to the square of its total momentum. Thus, $\Im\mathcal{A}$ is not zero only if the following constraint is satisfied:
\begin{equation}\label{eq:constraint}
\epsilon\left(\mathbf{k}, k_x\right)-\epsilon\left(\mathbf{k}+\mathbf{Q}, k_{1 x}\right)=0 \rightarrow k_x^2-k_{1 x}^2 \approx 2 k_z Q
\end{equation}
Then the square root of occupation numbers in eq. (\ref{eq: Im A}) can be roughly estimated as 1 and the product $V_1 V_2^*$ as

\begin{equation}\label{eq:V-1V-s-est}
V_1 V_2^* \approx \frac{512 \pi^4 \mu_B^3 M}{81 A d} \frac{k_x^2 k_x^{\prime 2}}{\left(k_z Q d\right)^4} \cos \frac{\left(k_x-k_{1 x}\right) d}{2} \cos \frac{\left(k_x^{\prime}-k_{1 x}\right) d}{2}
\end{equation}
For $\Re \mathcal{A}$ the constraint (\ref{eq:constraint}) is invalid. However, the range of the variable $k_{1 x}$ far from $k$ and $k^{\prime}$ so that $\left|k-k_{1 x}\right|,\left|k^{\prime}-k_{1 x}\right| \gg$ $1 / d$, contributes negligibly small amount to the real part of integral (\Ref{eq:Compton}). Indeed, in this range of the variable $k_{1 x}$ the vertices $V_1, V_2$ decrease as $1 /(k d)^2$ and strongly oscillate. Thus, in the region of $k_{1 x}$ substantial for the integral (\Ref{eq:Compton}) the inequalities $\left|k_x-k_{1 x-2 k_z Q}\right| \sim 1 / d$ and $\vert k_x^{\prime} - k_{1x}\vert\sim 1/d$ are satisfied. It means that $\Re \mathcal{A}$ has the same order of magnitude as $\Im \mathcal{A}$. Thus, we arrive at the estimate for the $|\mathcal{A}|^2$ :
\begin{equation}\label{eq:Im-A-square}
|\mathcal{A}|^2=\frac{32^2 \pi^8}{81^2 A^2 d^2} \frac{\mu_B^4}{k_T^2 Q^8 d^6 \ell^4}\left|\cos \frac{\left(k_x-k_x^{\prime}\right) d}{2}\right|^2 .
\end{equation}
Then
\begin{equation}\label{eq:tau-C-est}
\frac{1}{\tau_C} \approx \frac{2^7 \pi^8}{81^2} \frac{\mu_B^3}{k_T^3 Q^8 d^5 \ell^6 \hbar a^3 M}\left(n_{\mathbf{Q}}-n_{-\mathbf{Q}}\right) \approx 3 * 10^{-7} s^{-1} ,
\end{equation}
where $k_T=\min \left(\sqrt{\frac{k_B T}{2 \mu_B M}} \frac{1}{\ell}, \frac{\pi}{a}\right)$. In this calculation we assumed that $(n_{\mathbf{Q}} - n_{-\mathbf{Q}})\sim k_B T/(\mu_B H) \approx 10^5$ at $T=1000\,K$ and $H=1000\,Oe$.
Thus, the Compton relaxation time is $\tau_c=3.2 * 10^6 \mathrm{~s} \approx 38$ days. 

This estimate is rather rough and in reality may be by two-three -decimal orders smaller or larger. The second graph gives smaller value $\tau_C \approx 3-5 \,hours$. Still this time strongly exceeds the time of the experiment. Thus, we expect that equilibrium for inter-minimum transitions is not established and stationary or time dependent distribution of  magnons for this degree of freedom obeys the Boltzmann kinetic equation.

\section{Boltzmann kinetic equation}
The time evolution of the distribution function $n(\mathbf{k},t)$ is governed by the Boltzmann kinetic equation:

\begin{equation} \label{eq:Boltzmann}
\frac{dn}{dt}=\frac{\partial n}{\partial t} + \frac{\partial H}{\partial \mathbf{p}} \cdot \frac{\partial n}{\partial \mathbf{r}} + \mathbf{F} \cdot \frac{\partial n}{\partial \mathbf{p}} = I_{coll}
\end{equation}
Here the magnon states are characterized by their momenta. Since we assume that stationary state is homogeneous in real space, $ \frac{\partial n}{\partial \mathbf{r}} = 0$ and $\mathbf{F} = 0$, the terms with $\nabla_r n$ and $\nabla_p n$ in eq.(\ref{eq:Boltzmann}) vanish. In addition to the collision integral, one should also include into kinetic equation the pumping contribution. The number of magnons excited by pumping per unit time in total volume is equal to 

\begin{equation}\label{pump-number}
\frac{dN_m}{dt} =\frac{P}{\hbar\omega_0}
\end{equation}
where $P$ is absorbed power of pumping and $\omega_0$ is the frequency of parametric pumping generator. In stationary state this flow of pumped magnons must be compensated by the flow of disappearing magnons mostly due to the decay-merging processes. This gives equation of magnon balance:

\begin{equation}\label{eq:n-balance}
\frac{P}{\hbar\omega_0} = \frac{dN_{decay}}{dt} = V \int^{k_1} I_{coll}^{(3)}(\mathbf{k})\frac{d^3k}{2\pi^3},
\end {equation}
where V is the volume of the system; the limit $k_1=\mathrm{min}\left(\sqrt{k_BT/J}/l_d, \pi /a\right)$ shows the upper cutoff of integration over magnitude of $\mathbf{k}$. The notation $I_{coll}^{(3)}$ means that only part of the collision integral associated with decay-merging processes participates in the number of magnons balance.
This contribution to the Hamiltonian, as demonstrated in \cite{Gang-Li}, originates from the expansion of the Hamiltonian to the 3rd  order.

\begin{equation}\label{eq:3-order} 
\frac{\partial n(\mathbf{k},t)}{\partial t} = I(\mathbf{k}) = I_{pump}(\mathbf{k}) + I_{coll}^{(3)}(\mathbf{k}).
\end{equation}
For stationary states, $LHS=RHS=0$, eq.(\ref{eq:3-order}) is  simply the detailed balance condition:

\begin{equation}\label{eq:det-balance}
I_{pump}(\mathbf{k}) = -I^{(3)}_{coll}(\mathbf{k}).
\end{equation}
In principle this nonlinear equation for occupation numbers $n(\mathbf{k})$ can be solved approximately, but it is rather complicated because the collision integral includes not only magnon-magnon, but also magnon-phonon interactions. Here we apply another approach based on the conservation laws and variational principle.\\

We already employed the balance of total number of magnons that results in eq.(\ref{eq:n-balance}). The energy balance is given by:

\begin{equation}\label{eq:E-balance}
V \int_0^{k_1} \frac{d^3k}{(2\pi)^3} I_{coll}^{(3)}(\mathbf{k})\epsilon(\mathbf{k}) = \frac{dE}{dt} = P_{pump}
\end{equation}
We will look for a variational trial function for occupation numbers $n(\mathbf{k})$ in the form of step function depending on energy:

\begin{equation}\label{eq:1-step}
T(\epsilon) = \left\{ 
\begin{aligned} 
  T_1 & & \epsilon \leq \epsilon_1 \\ 
  T_0 & & \epsilon > \epsilon_1 
\end{aligned} \right.
\end{equation}
with two parameters $T_1$ and $\epsilon_1$, that can be found from the two balance conditions. The motivation for this Ansatz is the experimental observation made in the works \cite{Demokritov}, that low-energy magnons occupation numbers obey the Rayleigh-Jones law with effective temperature in the range about 1000 K. 

\section{Interaction terms in the ferromagnetic Hamiltonian}
The multiple interaction terms of the 3rd order found in the work \cite{Gang-Li} originate from the dipolar interaction. They are

\begin{equation}\label{eq:AR-3}
\begin{aligned}
& H_{d 3}=-\frac{2 \pi \gamma \sqrt{2 \gamma M}}{\sqrt{A}} \sum_{n_1, n_2, n_3} \sum_{\mathbf{q}_1, \mathbf{q}_2, \mathbf{q}_3} \delta_{\mathbf{q}_1-\mathbf{q}_2+\mathbf{q}_3} \\
& \times\left[\left(I_{d 3,---} \eta_{\mathbf{q}_1 n_1} \eta_{-\mathbf{q}_2 n_2} \eta_{\mathbf{q}_3 n_3}+I_{d 3,--+} \eta_{\mathbf{q}_1 n_1} \eta_{-\mathbf{q}_2 n_2} \eta_{-\mathbf{q}_3 n_3}^*\right.\right. \\
& \left.\left.+I_{d 3,+--} \eta_{-\mathbf{q}_1 n_1}^* \eta_{-\mathbf{q}_2 n_2} \eta_{\mathbf{q}_3 n_3}+I_{d 3,-+-} \eta_{\mathbf{q}_1 n_1} \eta_{\mathbf{q}_2 n_2}^* \eta_{\mathbf{q}_3 n_3}\right)+c . c .\right],
\end{aligned}
\end{equation}
where $\eta_{\mathbf{q},n}$ are coefficients of the Bogolyubov transformation in the wave-vector presentation; $n_{1,2,3}$ labels the transverse modes and $\mathbf{q}_{1,2,3}$ are the parallel wave vectors. At $q_y=0$ and $n_i=1$, the Bogoliubov transformation coefficients reads

\begin{equation}\label{eq:Bog-transverse}
\begin{aligned}
u & =a \cos \frac{\pi x}{d} \\
v & =c \cos \frac{\pi x}{d}
\end{aligned}
\end{equation}
The coefficients $I_{d3,\pm \pm \pm}$ can be evaluated as

\begin{equation}\label{eq:Bog-expl}
\begin{aligned}
I_{---} & =\frac{2 i a_1 c_2 c_3 d \pi^3 q_{3 z} \sinh d q_3}{4 \pi^4 q_3+5 d^2 \pi^2 q_3^2+d^4 q_3^5} \\
I_{--+} & =\frac{2 i a_1 c_2 c_3 d \pi^3 q_{3 z} \sinh d q_3}{4 \pi^4 q_3+5 d^2 \pi^2 q_3^2+d^4 q_3^5} \\
I_{+--} & =-\frac{2 i c_1 c_2 c_3 d \pi^3 q_{3 z} \sinh d q_3}{4 \pi^4 q_3+5 d^2 \pi^2 q_3^2+d^4 q_3^5} \\
I_{-+-}& =-\frac{2 i a_1 a_2 c_3 d \pi^3 q_{3 z} \sinh d q_3}{4 \pi^4 q_3+5 d^2 \pi^2 q_3^2+d^4 q_3^5}
\end{aligned}
\end{equation}
In these equations and till the end of this section we omit the upper index (3) in the interaction of three magnons. The reader should keep in mind that one of three momenta has the magnitude $\sim Q$ much smaller than two others which belong to thermal magnons.
For rough approximation at $qd\gg 1$ which is valid for thermal magnons, one can put $a_i,c_i \approx 1/\sqrt{d}$ and $\sinh dq_3=e^{dq_3}/2$. Neglecting also in denominators of eq.(\ref{eq:Bog-expl}) all terms except of the last one, we arrive at the result for the total interaction coefficient:

\begin{equation}\label{eq:I-fin}
I \approx \frac{2 i\pi^3 q_{3 z} e^{d q_3}}{d^{\,9/2} q_3^5}.
\end{equation}
Transition rate is then

\begin{equation}\label{eq:total-rate}
W=\frac{2 \pi}{\hbar}|I|^2 \delta\left[\omega_{\mathbf{k+q},n_3}-\omega_{\mathbf{k},n_1}-\omega_{\mathbf{q},n_2}\right]
\end{equation}

The total rate of occupation number of magnons with fixed wave vector $\mathbf{k}$ and number of branch $n$ in the Born approximation is
\begin{equation}\label{eq:Born}
\begin{aligned}
& \frac{d n_{\mathbf{k}, n}}{d t}=\sum_{n_1, n_2} \sum_{\mathbf{q}} W\left(\mathbf{k}, \mathbf{q}, n, n_1, n_2\right) \Big[n_{\mathbf{k}+\mathbf{q}, n_2} \times \\
& \left(n_{\mathbf{k}, n}+1\right)\left(n_{\mathbf{q}, n_1}+1\right)-\left(n_{(\mathbf{k}+\mathbf{q}, n_2}+1\right) n_{k, n} n_{q, n_1} \Big].
\end{aligned}
\end{equation}

\section{Born approximation}
The transition probabilities $W\left(\mathbf{k}, \mathbf{q}, n, n_1, n_2\right)$ can be calculated in the Born approximation. The matrix element of decay process of a thermal magnon into another thermal magnon and a magnon with wave vector $\pm \mathbf{Q}$,  $\langle \mathbf{k}\pm \mathbf{Q}\vert \mathbf{k}, \pm \mathbf{Q}\rangle$  is integral of the pair dipolar potential

\begin{equation}\label{eq:dip-pot-3}
U_{dip}^{(3)}=U_{dip}\left(\mathbf{r}_1-\mathbf{r}_2\right) + U_{dip}\left(\mathbf{r}_1-\mathbf{r}_0\right) + U_{dip}\left(\mathbf{r}_2-\mathbf{r}_0\right),
\end{equation}
where $\mathbf{r}_{1,2}$ are coordinates of the thermal magnons and $\mathbf{r}_0$ is coordinate of the magnon  with wave vector $\pm \mathbf{Q}$ with the normalized wave functions $\psi_{\mathbf{k}}\pm\mathbf{Q}(\mathbf{r}_1)^{*}$, $\psi_{\mathbf{k}}(\mathbf{r}_2)$, $\psi_{\pm \mathbf{Q}}(\mathbf{r}_0)$, respectively. Thus, the transition matrix element is integral of the dipolar potential 

\begin{equation}\label{eq:dip-pot}
U_{dip}(\mathbf{r}) = -\mu_B^2 \frac{r^2-3z^2}{r^3}
\end{equation}
between the states of thermal magnons $\mathbf{k}+\mathbf{Q}$ and $\mathbf{k}-\mathbf{Q}$ 
The differential cross-section of the potential provided by a low-energy magnon with momentum $q\sim Q$ can be written as

\begin{equation}
\frac{d\sigma}{d\Omega} = \frac{m^2\mu^4_B}{4\pi^2 \hbar^4}\cos^4\theta
\end{equation}
with $\theta = \arccos(q_z/q)$.
Thus, we can write the collision integral of the low-energy magnons as

\begin{equation}\label{eq:coll-int-Q--Q}
I(q) = - \int \frac{d^3k}{(2\pi)^3} v(\mathbf{k}) \frac{d\sigma}{d\Omega} \Big[ n_{k+q}(n_k+n_q+1) - n_k n_q \Big] \delta_{\epsilon_{\mathbf{k}+\mathbf{q}}, (\epsilon_{\mathbf{k}}+\epsilon_{\mathbf{q}})}
\end{equation}
where $v(\mathbf{k})= \frac{\partial\epsilon(\mathbf{k})}{\partial \mathbf{k}} = 2M\ell^2k$ provides the correct dimension, and can be interpreted as the in-flow current due to scattering.\\

From the energy conservation law $\epsilon_{\mathbf{k}+\mathbf{q}}-\epsilon_{\mathbf{k}} - \epsilon_{\mathbf{q}}=0$ one can find $k$ at a fixed $\mathbf{q}$:
\begin{equation}\label{eq:k-q}
k_{q,n}=\frac{\omega_{q,n}}{2\ell^2q\cos\theta},
\end{equation}
where in the interval $1\leq n\ll Qd$ the frequency $\omega_{\mathbf{q},n}$ is determined by 
\begin{equation}\label{eq:omega-q}
\omega_{q,n}= \gamma \sqrt{H^2+2HM\ell^2\left( q^2+\frac{Q^4}{q^2}\right)}.
\end{equation}
Substituting $k_q$ instead of $k$ in eq.(\ref{eq:coll-int-Q--Q}), we find:
\begin{equation}
I_{coll}^{(3)}(q) = -\frac{k_{q}^4}{16\pi^4}\gamma M\ell^2\frac{m^2\mu_B^4}{\hbar^4} \left[ n_{k_{q} + q}(n_{k_q} + n_q+1) - n_{k_q} n_q \right].
\end{equation}
This equation presents the decay rate of the low-energy magnon with momentum $\mathbf{q}$ through the $3^{rd}$ order dipolar interaction channel, which is the dominant part.
As discussed before, the quasi-equilibrium state is stationary as a result of the balance of both magnon number and energy between the parametric pumping and decay. 
\section{Rayleigh-Jones steps solution}
Kinetic equation for stationary state is
\begin{equation}\label{eq:kin-eq}
I_{coll}^{(3)}+I_{pump}=0,
\end{equation} 
where the rate of pumped magnons in the momentum space at pumping power $P$ is;
\begin{equation}\label{eq:I-pump}
I_{pump}(\mathbf{k}) = 2 \frac{P}{\epsilon_{\mathbf{k}}} \delta_{\epsilon_{\mathbf{k}},\frac{\omega_0}{2}}.
\end{equation}
The simplest variational solution of the kinetic equation is a step function of the form
\begin{equation}\label{eq:Ralyleigh-Jeans}
n(\epsilon) = 
\begin{cases}
\frac{T_1}{\epsilon - \mu} & \epsilon \leq \epsilon_1, \\
\frac{T}{\epsilon-\mu} & \epsilon_1 \leq \epsilon \leq \epsilon_{\mathrm{max}}.
\end{cases}
\end{equation}
In the previous section we already discussed the motivation of the Rayleigh-Jones distribution by experimental data. For numerical calculation we choose the pumping energy $\epsilon_{pump} = 1.5*\Delta$. 

The effective chemical potential can be taken as $\mu = k_B (T_1 - T$, which increases up to the ground state energy $\Delta = g\mu_B H$ when $T_1$ grows from $T$ to the critical temperature  $(T_1)_c = T_0 + \Delta/ k_B$. Together with a chosen pumping frequency $\epsilon_{pump}$, we obtain $P_{pump-critical} = 6\,W$ and $\epsilon_1 = \hbar 10\,GHz$, where $\epsilon_1$ lands reasonably between the low-energy region and thermal region.

\section{Maximum Entropy Production Principle (MEPP)}
In a slightly more sophisticated version, we consider the temperature that has two steps of the form
\begin{equation}\label{2-step}
T(k) = \left\{ 
\begin{aligned} 
  T_1 & & \epsilon \leq \epsilon_1 \\ 
  T_2 & & \epsilon_1 <\epsilon \leq \epsilon_2 \\ 
  T_0 & & \epsilon > \epsilon_2 
\end{aligned} \right.
\end{equation}

Such a function has 4 parameters, which cannot be found solely from the balance of number of magnons and their energy. Thus we need to employ the maximum entropy production principle.\\

Recent developments in nonequilibrium statistical physics have substantially deepened the theoretical foundation of the Minimum Entropy Production Principle (MEPP)
and broadened its range of applicability. Dechant and co-workers have provided a rigorous geometric formulation of MEPP for continuous-time Markov processes, revealing its connection to the Wasserstein distance and optimal transport theory, and demonstrating
that minimum-dissipation trajectories naturally emerge under appropriate dynamical constraints \cite{Dechant1, Dechant2, Yoshimura, VanVu}. Subsequent works have extended these concepts to general nonlinear dynamics and to coarse-grained macroscopic systems, establishing a unifying thermodynamic framework that encompasses both microscopic stochastic processes and emergent macroscopic behavior \cite{Yoshimura, Falasco}. Moreover, explicit analysis of resetting and other far-from-equilibrium processes have shown that entropy production can be consistently characterized and minimized even in strongly driven regimes \cite{Mori}. The conceptual reach of the MEPP has also expanded beyond traditional physics, finding applications in information processing and machine learning, where it guides the optimization of graph-based neural architectures \cite{Yang}.These studies affirm the MEPP as a robust and versatile theoretical tool for quantifying non-equilibrium dissipation. On this basis, applying the MEPP to describe the non-equilibrium state of magnon condensates is both physically justified and conceptually consistent with the modern understanding of entropy production in driven systems.

The entropy of the magnons is 
\begin{equation}\label{eq:entropy}
  S = \int_\Delta^{\epsilon_{max}} n(\epsilon)ln[n(\epsilon)]\nu(\epsilon)d\epsilon  
\end{equation}

The entropy production can thus be written as 
\begin{equation}\label{eq:entropy-prod}
  EP=\frac{dS}{dt} = \int \frac{dn(\epsilon)}{dt} \big[ln[n(\epsilon)]+1\big] \nu(\epsilon)d\epsilon 
\end{equation} 

Here we consider stationary state. Therefore the derivatives are literally zero. However, due to pumping, the magnons move to higher energy state with local 
rate of $n(\epsilon)/\tau(\epsilon)$, where $1/ \tau(\epsilon)$ is the probability of transition per unit time averaged over the surface of constant energy in the momentum space.
\begin{equation}\label{eq:ep-average}
  EP = \int \frac{n(\epsilon)}{\tau(\epsilon)} \big[ln[n(\epsilon)]+1\big] \nu(\epsilon)d\epsilon 
\end{equation}
with

\begin{equation}
  \frac{1}{\tau(\epsilon)}=\frac{\int_{\epsilon} d\mathbf{k}I(\mathbf{k})} {\int_{\epsilon} d\mathbf{k}},
\end{equation}
where the symbol $\int_{\epsilon}$ means the integration inside the volume in momentum space limited by the surface $\epsilon(\mathbf{k})=\epsilon$.
The 4 parameters $(T_1,T_2,\epsilon_1,\epsilon_2)$ are estimated by maximizing the value of $EP$. 

\begin{figure}[h]
    \centering
    \includegraphics[width=0.5\textwidth]{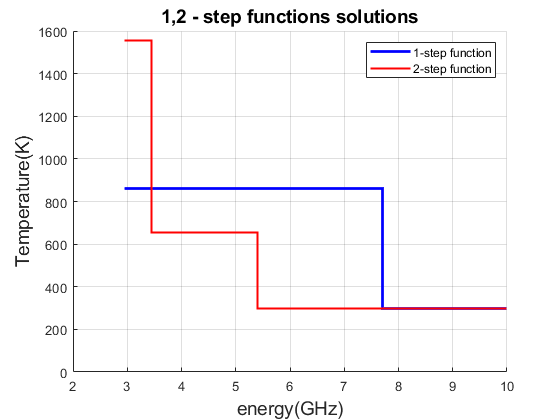}
    \caption{The numerical results for the 1-step and 2-step function models. !-step function is shown by solid lines, the two-step function by thinner line . They demonstrate that the low energy magnons have higher temperature than the thermal magnons. We expect that the n-step  diagram will asymptotically approach a continuous steady state curve, which characterizes the actual distribution of magnons accumulating near the ground state under pumping.}
    \label{step_sol}
\end{figure}

Redefining $R(\epsilon)=\frac{\nu(\epsilon)}{\tau(\epsilon)}$, together with the magnon number and energy balance, we have
\begin{equation}\label{EP}
\begin{aligned}
& EP = \int_\Delta^{\epsilon_2} n(\epsilon) \big[ln[n(\epsilon)]+1\big] R(\epsilon) d\epsilon\,,\\ 
& \dot{N} = \int_\Delta^{\epsilon_2} n(\epsilon) R(\epsilon) d\epsilon\,,\\ 
& \dot{E} = \int_\Delta^{\epsilon_2} n(\epsilon) R(\epsilon) \epsilon d\epsilon\,.
\end{aligned}
\end{equation}
We can use the Rayleigh-Jones occupation number $n(\epsilon)=\frac{k_B T}{\epsilon}$ for  for the long-wave magnons.  For convenience, we define following integrats
\begin{equation}
\begin{aligned}
& A(\epsilon) = \int R(\epsilon) d\epsilon\,,\\ 
& B(\epsilon) = \int \frac{1}{\epsilon} R(\epsilon) d\epsilon\,,\\  
& C(\epsilon) = \int \frac{ln(\epsilon)}{\epsilon} R(\epsilon) d\epsilon\,.
\end{aligned}
\end{equation}
The limits of integrations are the same as in eq.(\ref{EP}).
Then for the 2-step function in eq.(\ref{2-step}), we have
\begin{equation}
\begin{aligned}
 EP &= \dot{N}+T_1\,ln(T_1)\left[B(\epsilon_1)-B(\Delta)\right]+T_2\,ln(T_2)\left[B(\epsilon_2)-B(\epsilon_1)\right]\\
 &-T_1\left[C(\epsilon_1)-C(\Delta)\right]-T_2\left[C(\epsilon_2)-C(\epsilon_1)\right]\,,\\ 
 \dot{N} &= T_1\left[B(\epsilon_1)-B(\Delta)\right]+T_2\left[B(\epsilon_2)-B(\epsilon_1)\right]\,,\\ 
 \dot{E} &= T_1\left[A(\epsilon_1)-A(\Delta)\right]+T_2\left[A(\epsilon_2)-A(\epsilon_1)\right]\,.
\end{aligned}
\end{equation}
We then apply the Lagrange multiplier method to find the $T_1, T_2, \epsilon_1, \epsilon_2$ parameters that maximize entropy production $EP$. Define the Lagrangian as
\begin{equation}
\mathcal{L}=EP-\lambda_N\dot{N}-\lambda_E\dot{E}\,.
\end{equation}
Then we set the partial derivatives to zero
\begin{equation}
\begin{aligned}
\frac{\partial\mathcal{L}}{\partial T_1}=0,\,
\frac{\partial\mathcal{L}}{\partial T_2}=0,\,
\frac{\partial\mathcal{L}}{\partial \epsilon_1}=0,\,
\frac{\partial\mathcal{L}}{\partial \epsilon_2}=0\,.
\end{aligned}
\end{equation}
As shown in Fig.\ref{step_sol}, this gives the results: $T_1=1556\,K$, $\epsilon_1=3.45\,GHz$, $T_2=654\,K$, $\epsilon_2=5.4\,GHz$, as shown along with the 1-step function result in Fig.\ref{step_sol}. 
Again, the actually effective temperature should be $T_1=1856\,K$ and $T_2=954\,K$, to compensate for the decay rates without pumping.  The higher peak at the low-energy region appears as a result of maximizing the entropy production, since extreme points of the $n\,log\,n$ entropy-like quantity will have more distribution at lower weight regions. In some experiments, they observed that the effective temperature for low energy magnons was about $1000\,K$. i.e., about twice bigger than the Curie point.

\section*{Acknowledgments}
Acknowledgments statement.

\section*{Author Contributions}
H. Yang: Solution of the kinetic equation for one-step and two-step variational distribution functions for magnons; numerical calculation of transition probabilities. G. Li: Calculation of quasi-equilibrium distribution function of magnons and their three-particle and four-particle interactions; participation in the experiment that proved repulsive interaction between magnons. H. Jia: Numerical calculations in an early version of the slow relaxation time. A. Abanov: Proposal of the step-like distribution function. V. Pokrovsky: General idea of inter-minima relaxation as a slow process and its realization through Compton and four-vertex scattering.

\section*{Funding Statement}
This work was financially supported by Dr. Valery Pokrovsky under the auspices of 02-512992 William R Thurman \textquoteright58 Chair Physics.

\section*{Data Availability}
Data will be made available on request.

\end{document}